# A Physics-Guided Neural Framework for Rheology Measurement from Dynamical Laser Speckles

Tianliang Wang[1], Thomas Goudoulas[1], Ehsan Fattahi[1], Dominik Geier[1], Yiyuan Yang[2], Ivan Ezhov[3], Yixiao Liu[4], Yi Li[5], Martin Booth[6], Thomas Becker[1,*]

[1]*Group of BioPAT and Digitalization, TUM School of Life Sciences, Technical University of Munich, Freising 85354, Germany*

[2]*Department of Computer Science, University of Oxford, Oxford OX1 3QD, UK*

[3]*TUM University Hospital, Technical University of Munich, Munich 80333, Germany*

[4]*Department of Urology, Peking University Third Hospital, Beijing 100191, China*

[5]*College of Optical and Electronic Technology, China Jiliang University, Hangzhou 310018, Zhejiang, China*

[6]*Department of Engineering Science, University of Oxford, Oxford OX1 3PJ, UK*

[*]*email: tb@tum.de*

**Abstract**

Critical breakthroughs in the area of biomedicine and materials science increasingly depend on rapid, non-contact methods for viscoelastic characterization. Laser Speckle Rheology (LSR) is positioned to meet this demand, effectively circumventing the speed and invasiveness bottlenecks inherent to traditional mechanical rheometry. However, its application in turbid fluids is severely constrained by multiple scattering, where standard physical inversions rely heavily on precise, sample-specific optical transport parameters that are difficult to measure in situ. To overcome this barrier, we propose a physics-guided deep learning framework that infers a Maxwell relaxation spectrum from the intensity autocorrelation $g_2(t)$ and speckle-intensity histogram statistics. The resulting spectrum is then propagated through a Maxwell forward model to predict $G'(\omega)$ and $G''(\omega)$ under physics-consistency constraints. Quantitatively, the framework achieves $\text{RMSE}_{\log}$ as low as 0.009 against reference rheometry and generalizes to previously unseen scattering conditions, preserving physically plausible frequency dependence and $G'(\omega)$–$G''(\omega)$ phase behavior. It reduces reliance on optical transport parameters that are hard to determine in situ and returns an interpretable generalized Maxwell relaxation spectrum, improving the practicality of LSR in turbid media.

## Introduction

The storage modulus ($G'(\omega)$) and loss modulus ($G''(\omega)$) are critical parameters used to characterize the viscoelastic behavior of complex fluids and soft-matter systems. This dynamic information of the sample they provided spans across biomedical engineering, food science, industry, and biology, which makes their measurement essential for understanding material performance and guiding practical applications [1-8]. Initially, $G'(\omega)$ and $G''(\omega)$ were measured using mechanical setups such as rotational rheometers and dynamic shear rheometers [9]. These devices evaluate the mechanical properties of a sample by applying continuous rotational shear forces, as well as periodic shear stress or strain, and then meticulously measure the sample's response at a range of angular frequencies [10-14]. However, this type of method is invasive, time-consuming, requires large sample volumes, relies on expensive, bulky instrumentation, and is only usable offline [12, 14-17]. Their applicability in many practical and in situ scenarios is limited. To address these limitations, alternative techniques such as dynamic light scattering (DLS) and diffusing wave spectroscopy (DWS) have been developed. DLS measures sample dynamics and gives an answer about diffusional processes by analyzing the time correlation of scattered light intensity, providing fast, simple, and highly sensitive measurements [18]. Nevertheless, its applicability to concentrated or polydisperse samples is limited (for example, due to local sample fracture or high normal forces), often requiring dilution, which can then contaminate the samples [19]. DWS, by contrast, extends measurements to turbid or concentrated systems, partially overcoming DLS's shortcomings [20, 21]. However, its accuracy strongly depends on the correct determination of optical transport parameters, such as transport mean free path, and it is sensitive to absorption and sample geometry. Small mismatches in these parameters can lead to systematic bias in the recovered correlation functions and, consequently, in the derived MSD and viscoelastic moduli [22-25].

An all-optical technique called laser speckle rheology (LSR) was then introduced to address the limitations of DLS and DWS. Traditional LSR methods typically involve capturing the speckle intensity autocorrelation $g_2(t)$, then inferring the particle mean square displacement (MSD) by fitting $g_2(t)$ to models such as the stretching exponent. Finally, the material's complex moduli, $G^*(\omega)=G'(\omega)+ iG''(\omega)$ are calculated via the generalized Stokes-Einstein relation (GSER) [26-28]. This approach offers significant advantages: it enables fast, non-invasive measurements, particularly at high frequencies beyond the reach of traditional rheometers, and is applicable to optically dense or turbid systems that are difficult to handle with other methods. However, the implementation of this principle as a reliable and widely applicable tool is challenging, representing a significant issue for the industry. First, multiple scattering in dense or strongly scattering media can severely distort the $g_2(t)$ signal, invalidating simple scattering models and complicating interpretation [24, 26, 27]. Second, inverting $g_2(t)$ to obtain the MSD and subsequently $G^*(\omega)$, the core computational step, is an inherently ill-posed problem. Classical inversion approaches often fit the measured $g_2(t)$ using simplified decay models [28]. Yet, not all experimental $g_2(t)$ curves can be accurately represented by such models, and in many cases, substantial deviations remain between the fitted and measured data, causing them to deviate from the assumed model shape and often yielding unstable or physically ambiguous MSD results [27]. Crucially, any errors or instabilities in the inferred MSD are further amplified when converted to the frequency domain to obtain $G^*(\omega)$ via the GSER, leading to significant inaccuracies in the final viscoelastic spectrum [25]. Furthermore, subtle changes in experimental conditions, such as particle size, concentration, solvent properties, or even instrument optics, can significantly alter the apparent $g_2(t)$ shape due to confounding optical effects [29]. While advanced simulation tools like polarization-sensitive correlated transport Monte Carlo ray tracing (PSCT-MCRT) are designed to account for these complex photon paths [27], their practical utility is constrained by a dependency on sample-specific recalibration. Each new formulation requires the model to be fed with prior knowledge of its precise optical transport parameters. As a result, even complex simulation models must be recalibrated regularly to reflect the sample-specific parameters and attenuation characteristics. This severely limits their robustness and portability across different formulations, hindering their use as reliable "one-click" tools for rapid formulation screening or in-line process monitoring.

These central limitations highlight the clear need for a new strategy that circumvents the inherent instabilities of direct inversion. In this study, we propose a physics-guided neural network architecture to reconstruct viscoelastic spectra $G'(\omega)$ and $G''(\omega)$ directly from speckle intensity time series. Crucially, rather than treating the problem as a purely data-driven regression task, we reformulate the ill-posed inversion as a physics-constrained parameter estimation problem. The network is trained to infer physically admissible relaxation spectra ($H(\tau)$) that are consistent with both the observed speckle dynamics and fundamental viscoelastic constraints, thereby embedding prior physical structure directly into the learning process.

This formulation enables the model to suppress noise amplification inherent to direct numerical inversion and to distinguish physically meaningful decay behavior from measurement-induced distortions, while avoiding reliance on unstable intermediate quantities. Building on this physics-guided formulation, the proposed framework integrates multi-scale physical representations and constrained forward modeling to ensure Maxwell consistency and causal behavior. To further decouple optical variations due to particle size, concentration, and scattering conditions from the intrinsic Brownian dynamics, we incorporate auxiliary statistical information to enhance robustness across diverse experimental conditions. Extensive experiments with previously model-unseen particle sizes, concentrations, and solvents demonstrate that the proposed approach maintains physically meaningful relaxation spectra and achieves accurate reconstruction of $G'(\omega)$ and $G''(\omega)$ even under multi-scattering and high-noise conditions.

**Physics constrained network design**

Our primary innovation is a physics-constrained network architecture that embeds constitutive viscoelastic structure into the learned representation. The goal is to let the network learn the deterministic mapping from the measured $g_2(t)$, which remains a nonlinear mapping, and the histogram to the relaxation spectrum $H(\tau)$, while maintaining the physical constraints of non-negativity, smoothness, and causal consistency. An overview of how these physical constraints are translated into a concrete processing pipeline is provided in Figure 1. The $g_2(t)$ is first processed by a Decay Analyzer to extract multi-scale decay features, which are used to form an initial relaxation-spectrum representation (generalized Maxwell modes $\{G_i, \tau_i\}$ on the $\log \tau$ axis. These $g_2(t)$ features are then combined with Histogram features through an attention fusion module, and a Decoder refines the mode parameters to produce the final relaxation modes $\{G_i, \tau_i\}$. Finally, a Physics Forward step maps $\{G_i, \tau_i\}$ to the frequency-domain outputs $G'(\omega)$ and

$G''(\omega)$.

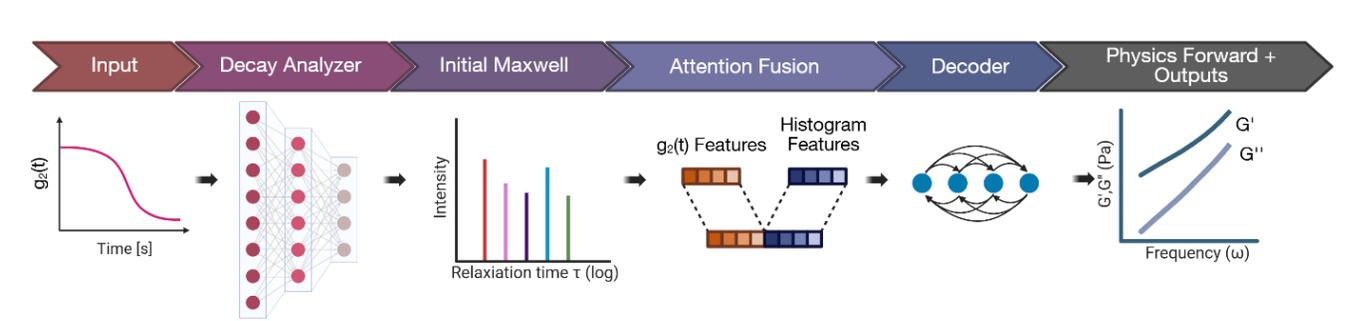

Figure 1. Schematic of the proposed physics-constrained network. The input autocorrelation $g_2(t)$ is encoded by a decay analyzer to form an initial generalized Maxwell spectrum $\{G_i, \tau_i\}$. Histogram-derived static features are fused with the $g_2(t)$ features via an attention module, and a decoder refines the relaxation modes. The final deterministic spectrum is mapped to frequency-domain outputs $G'(\omega)$ and $G''(\omega)$.

The detailed structure is shown in Figure 2. The analyzer converts the raw $g_2(t)$ into multi-scale [30], physically interpretable descriptors that isolate characteristic times and amplitudes to seed an initial Maxwell spectrum. The process begins with data enhancement (Figure 2, I), where the analyzer module first enhances the raw autocorrelation curve $g_2(t)$ by generating four complementary feature channels. These channels include: (i) the original decay curve, (ii) its first-order derivative, (iii) its logarithmic-time gradient, and (iv) its logarithmic-time curvature. By highlighting different aspects of the decay shape, such as rapid decorrelation intervals and inflection points, these enhanced features facilitate the subsequent extraction of multi-scale relaxation times $\tau$ and amplitudes by parallel convolutional layers, forming an initial estimate for the Maxwell spectrum.

Next, in the multi-scale mapping stage (Figure 2, II), the four-way enhanced signal is processed. We construct *S* relaxation times on a log-scale within $[\tau_{\min}, \tau_{\max}]$, where $\tau_{\min}$ and $\tau_{\max}$ are set according to the measurement time window and the material's viscoelastic properties. Each $\tau$ corresponds to a one-dimensional convolution branch processing the four input channels in parallel. Within each branch, the processing involves an initial physically-initialized convolution layer (Phys-init Conv1d), followed by a standard block of convolution, batch normalization (BN), and ReLU activation (Conv / BN / ReLU). The Phys-init Conv1d layer physically initializes kernels based on exponential decay ($e^{-t/\tau}$) for the original $g_2(t)$ channel and its derivative ($(1/\tau)e^{-t/\tau}$) for the first-order derivative channel, motivated by the generalized Maxwell model, while other channels use conventional initialization. The outputs across all scales are concatenated (Concat over scales) and then reweighted using a lightweight squeeze-and-excitation (SE) attention block (SE attention), effectively selecting the most informative time scales [31]. Following this, Initial Parameter Extraction occurs (Figure 2, III). The scale-reweighted features are passed through two parallel heads: an amplitude head that outputs non-negative amplitudes $A_s$, and a time head that slightly adjusts a fixed logarithmic grid of candidate times to obtain the final time constants $\tau_s$. This output matrix $\{A_s, \tau_s\}$ should be interpreted as an observational (overcomplete) exponential representation of $g_2(t)$, i.e., $\hat{g}_2(t) = \sum_{s=1}^{S} A_s e^{-t/\tau_s}$. Because multiple exponentials with nearby $\tau_s$ can explain the same finite-time-window decay, these components may be numerous and overlapping, and therefore do not necessarily constitute well-separated physical Maxwell modes.

Finally, the Maxwell spectrum projection stage (Figure 2, IV) outputs the Maxwell spectrum. Instead of projecting onto a static $\tau$-grid, we implement an adaptive discretization via a

location–scale transformation [32]. Specifically, a sub-network estimates the log-center and log-spread of the relaxation-time distribution from the multi-scale features together with the preliminary decay parameters ($\{A_s, \tau_s\}$)[33]. We use these moments to dynamically shift and scale a base template, thereby parameterizing the candidate relaxation times $\tau_i$. The resulting candidates are clamped in log-space to a physically admissible range, yielding an implicit parametric binning of the Maxwell modes. The corresponding mode strengths are predicted with softplus to enforce non-negativity and combined with softmax-normalized mode weights (summing to one) to yield the final spectrum $\{G_i, \tau_i\}$ [34]. This parametric estimation restricts the derived modes to a physically admissible set (non-negative $G_i$, bounded $\tau_i$, and normalized weights), while improving statistical stability.

To further refine these modes before computing $G'(\omega)$ and $G''(\omega)$, the dynamic features extracted from $g_2(t)$ serve as the query, while histogram-derived static features provide key–value pairs in a cross-attention fusion block [35]. This embeds concentration- and particle-size–related corrections without altering the intrinsic spectral shape. The fused representation is then passed through a lightweight self-attention decoder [36], enabling different relaxation components to interact under a shared context and make only small, physically feasible adjustments to their amplitudes and time constants. The final Maxwell spectrum $\{G_i, \tau_i\}$ is subsequently used in the standard Maxwell forward model to compute $G'(\omega)$ and $G''(\omega)$ [37]. A detailed description of the network configuration is provided in the Supplementary Information.

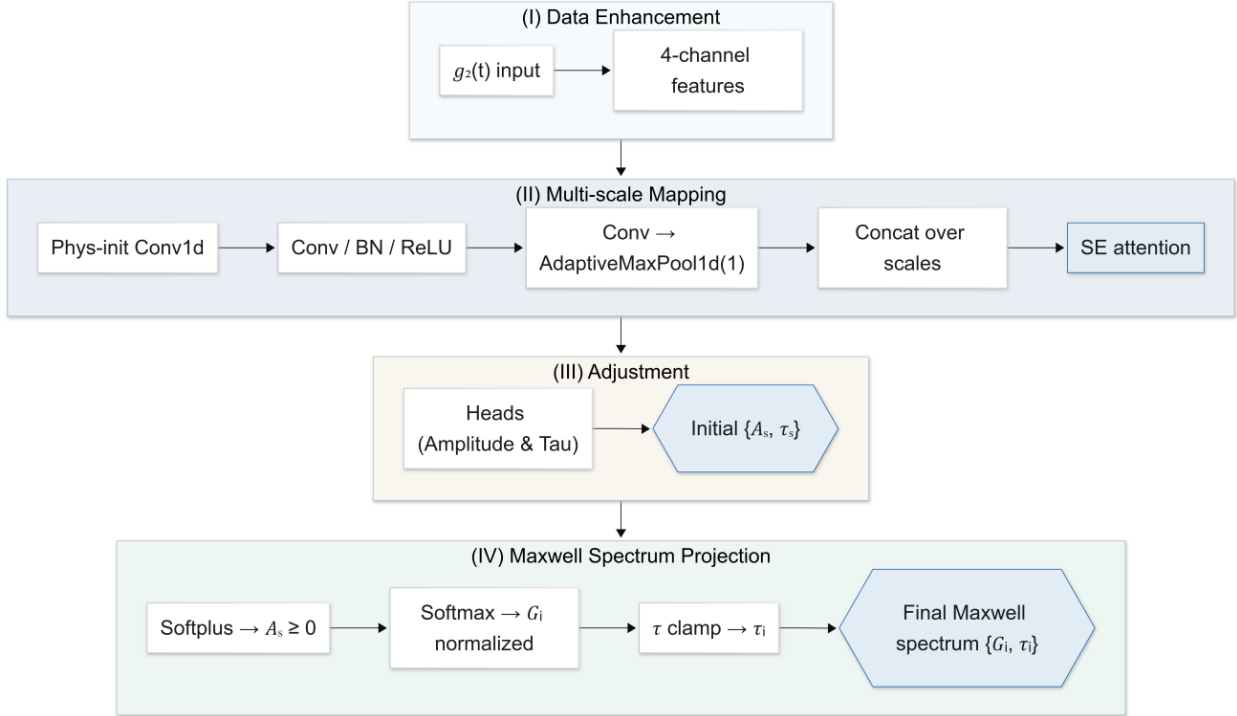

Figure 2 Flowchart illustrating the main processing stages of the physics-guided neural network. The architecture includes (I) data enhancement, (II) multi-scale mapping with physical initialization and attention, (III) initial parameter extraction, and (IV) projection onto a physically constrained Maxwell spectrum $\{G_i, \tau_i\}$.

While the architecture imposes structural physical constraints on the representation, we further enforce quantitative agreement with measured spectra by adding complementary physics-based loss terms during training.

**Loss function design**

To guide the optimization toward physically meaningful solutions, we introduce a set of physics-guided loss functions that explicitly couple the network predictions to experimentally observable quantities. These losses play a complementary role to the architectural constraints, stabilizing training and suppressing physically implausible solutions that may still satisfy the network structure alone. First, a frequency-domain physics loss ($L_{\text{phy}}$), comparing predicted and experimental $G'$ and $G''$, ensures that the final output directly corresponds to the experimentally measurable rheological spectrum. It can be seen from Eq. (1) that the loss is evaluated on a unified logarithmically spaced frequency grid $\{\omega_k\}_{k=1}^{N_\omega}$, where $k$ indexes the discrete frequency samples and $N_\omega$ denotes the total number of frequency points.

To account for the fact that $G'(\omega)$ and $G''(\omega)$ may differ by several orders of magnitude, the modulus mismatch is computed in the log-modulus domain, so that the loss measures relative rather than absolute errors and avoids domination by the larger component. In addition, a penalty on $\tan \delta(\omega) = G''(\omega)/G'(\omega)$ is included to constrain the phase relationship between elastic and viscous responses.

$$L_{phy} = \| log_{10}(G'_{pred} + \epsilon) - log_{10}(G'_{ref} + \epsilon) \|_1 + \| log_{10}(G''_{pred} + \epsilon) - log_{10}(G''_{ref} + \epsilon) \|_1 +$$
$$\left( \| s'_{pred} - s'_{ref} \|_2^2 + \| s''_{pred} - s''_{ref} \|_2^2 \right) + \| log_{10}(tan\delta_{pred} + \epsilon) - log_{10}(ta\delta_{ref} + \epsilon) \|_2^2 \quad (1)$$

The slope terms $s(\omega) = d\log G(\omega)/d\log \omega$ enforce the physically meaningful local power-law scaling behavior of the viscoelastic spectrum, which characterizes the frequency-dependent relaxation mechanisms of the material.

Second, on the $g_2(t)$ path, we use a reconstruction loss, $L_{g2}$, requiring that the multi-scale components extracted by the analyzer can be recombined into the original $g_2(t)$ curve, thereby preventing feature learning from becoming too abstract. As shown in Eq. (2), $\widehat{g_2}(t_n)$ is the reconstructed curve, the $g_2(t_n)$ is the input curve (original $g_2(t)$), N is the number of speckle pattern acquisition time points. $w_n$ here is the weight value, and $\frac{1}{N}\sum_n w_n = 1$ means the weights must be 1 after normalization to emphasize the early time period, thus avoiding over-abstraction of feature learning.

$$L_{g_2} = \frac{1}{N}\sum_{n=1}^{N} w_n \left(\widehat{g_2}(t_n) - g_2(t_n)\right)^2, w_n > 0, \frac{1}{N}\sum_n w_n = 1 \tag{2}$$

Meanwhile, to prevent all modes from shrinking to a single time scale (i.e., mode collapse), we design a $\tau$ separation constraint ($L_{scp}$) and an information entropy constraint ($L_{ent}$) to encourage the modes to be distributed within a reasonable range, as Eqs. (3)(4):

$$L_{scp} = \frac{1}{M-1}\sum_{i=1}^{M-1} max\{0, \delta - \left(\log_{10}\tau_{(i+1)}^{mx} - \log_{10}\tau_{(i)}^{mx}\right)\} \tag{3}$$

$$L_{ent} = |H(w) - H^\star|, \ H(w) = -\sum_{i=1}^{M} w_i \log w_i \tag{4}$$

where $M$ is the number of Maxwell modes and $\tau_i^{mx}$ are their logarithmically ordered relaxation times. The threshold $\delta$ defines the minimum spacing between adjacent modes to avoid collapse. Besides, $w_i$ denotes the normalized modal weights ($w_i > 0$, $\sum_i w_i = 1$), and $H(w) = -\sum_i w_i \log w_i$ is their entropy. The reference $H^\star$ controls dispersion, keeping the spectrum neither too concentrated nor too uniform. The final loss can be written as $L = 5L_{phy} + L_{g_2} + L_{scp} + L_{ent}$, where the physics loss is weighted by 5 (see supplementary material for more information). Together, the architecture and loss functions encourage solutions that remain grounded in physically interpretable relaxation behavior. In the next section, we briefly clarify the physical basis of this relaxation representation.

**Maxwell Model as a Physical Constraint in the Network**

In the following, we examine the physical basis of this intermediate description and explain why it provides a suitable link between speckle-derived decay features and macroscopic viscoelastic response. In this work, we introduce a strategy that bypasses the unstable GSER inversion. Our approach adopts the non-negative Maxwell relaxation spectrum, $H(\tau)$, as a core physical intermediate variable, yielding a compact, interpretable fingerprint (relaxation times and strengths).

In this formulation, the constitutive physical law is explicitly embedded into the network's forward pass. Rather than treating $H(\tau)$ as a purely abstract latent variable, it is used as a physically rigorous representation of the material's relaxation states. The subsequent mapping from $H(\tau)$ to $G'(\omega)$ and $G''(\omega)$ is implemented as a fixed, non-trainable summation layer defined by the generalized Maxwell equation. As a result, the network is restricted to predicting only the underlying relaxation spectrum, while the final viscoelastic moduli are obtained strictly through deterministic physical laws. This design imposes a hard physical constraint on the model output, and as a direct consequence, guarantees that the predicted spectra satisfy causality and the Kramers–Kronig relations [37], regardless of data noise.

Here, $H(\tau) \geq 0$ describes the distribution of relaxation intensities over characteristic relaxation times ($\tau$) within the material, providing a physically interpretable link between micro- and macro-scale behavior. The $H(\tau)$ rationale derives from its well-posed physical and mathematical properties. Physically, for linear passive viscoelastic materials, the field autocorrelation function $g_1(t)$ can be expressed as the Laplace transform of $H(\tau)$, as guaranteed by Bernstein's theorem [28, 35], shown in Eq. (5):

$$|g_1(t)|^2 = \int_0^\infty \exp(-t/\tau)\, H(\tau)\, d\ln\tau \qquad (5)$$

This formulation retains $H(\tau)$ as a well-posed, non-negative quantity, while MSD reconstruction involves differentiation and kernel inversion steps that are more sensitive to noise.

Mathematically, the mapping from $H(\tau)$ to the frequency-domain viscoelastic moduli, $G'(\omega)$ and $G''(\omega)$, is a stable Stieltjes linear integral transform [28]:

$$G'(\omega) = \int_0^\infty H(\tau) \frac{\omega^2 \tau^2}{1+\omega^2\tau^2} \, d\ln\tau \tag{6}$$

$$G''(\omega) = \int_0^\infty H(\tau) \frac{\omega\tau}{1+\omega^2\tau^2} \, d\ln\tau \tag{7}$$

$H(\tau)$ is a spectrum that helps us solve the problem of finding $G'(\omega)/G''(\omega)$ from $g_2(t)$. It simplifies the problem by breaking it down into a two-step process: first, find $H(\tau)$, then determine $G'(\omega)/G''(\omega)$ by integrating it in a steady manner.

However, relying solely on the measured $g_2(t)$ is insufficient for a reliable inference of $H(\tau)$. The measured intensity autocorrelation $g_2(t)$ relates to the physical $g_1(t)$ via the Siegert relation [38], $g_2(t) - 1 = \beta|g_1(t)|^2 = exp\{-2\gamma[k^2\langle\Delta r^2(t)\rangle]^\xi\}$. Here, $k$ is the wave number, while $\gamma$ and $\zeta$ are experimental constants reflecting the static optical properties of the sample. These parameters create ambiguity and make it difficult to recover a unique $H(\tau)$ from the $g_2(t)$ curve alone [16]. This is precisely where the intensity histogram is introduced as an independent constraint. It provides static, sample-dependent information that allows the network to effectively decouple the optical effects (encoded in $\gamma$ and $\zeta$) from the intrinsic Brownian dynamics (encoded in $g_1(t)$). By using both $g_2(t)$ and the histogram, $H(\tau)$ can be robustly inferred. Once a reliable $H(\tau)$ is obtained, $G'(\omega)$ and $G''(\omega)$ can be accurately and deterministically calculated through the stable linear integrals in Eqs. (6) and (7). Experimentally, this requires a setup that records the full speckle field with sufficient temporal resolution so that both $g_2(t)$ and the intensity histogram can be computed from the same acquisition. We next describe the experimental configuration.

## Experiments

**Experimental setup**

The underlying principle of the LSR is to analyze the intensity of the backscattered light from the sample to obtain the $G'(\omega)$ and $(G''(\omega)$. The experimental setup is depicted in Figure 3. A He-Ne laser (R-30995, 17 mW, Newport Spectra-Physics GmbH, Darmstadt, Germany) serves as the primary light source. After passing through the first polarizer (P1), the laser is split by a beam splitter (CCM1-BS013, Thorlabs, Newton, New Jersey, USA). One portion of the laser beam is directed to the optical trap, while the other 50% is used to illuminate the sample. The backscattered light from the sample is collected as the signal for intensity decay analysis. The backscattered light is then reflected by the beam splitter and collected by a high-speed camera (acA640-750uc, Basler AG, Ahrensburg, Germany) through a crossed polarizer (P2) and lens system 2, a 5× infinity-corrected objective (stock #15-941, Edmund Optics, Barrington, New Jersey, USA), and a camera tube lens (stock #18-504, Edmund Optics, Barrington, New Jersey, USA)). The crossed-polarization configuration suppresses co-polarized specular reflections and illumination leakage, improving speckle contrast for correlation analysis [28]. Given that $G'(\omega)$ and $G''(\omega)$ are highly sensitive to temperature, the environmental temperature is controlled consistently at 25 °C. Additionally, to mitigate the effects of ambient conditions, the entire setup is enclosed in an optical enclosure. This configuration allows both the temporal autocorrelation and the intensity statistics to be extracted from the same measurement.

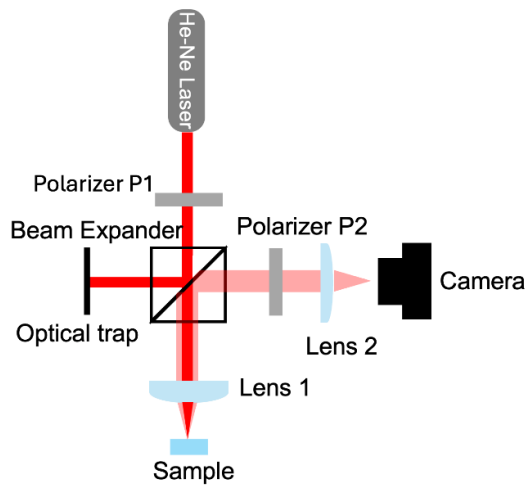

Figure 3 Schematic diagram of the laser speckle rheology experimental setup. After passing through a polarizer (P1) and a beam expander, the He-Ne laser enters the optical trap and is focused on the sample. The scattered light (lighter red colored) passes through lenses (Lens 2) and a polarizer (P2) and is captured by a camera to analyze the speckle pattern.

**Sample preparation**

The speckle pattern samples were prepared by mixing glycerol (≥99.5% purity, AnalaR® NORMAPUR® ACS analytical reagent, redistilled, VWR Chemicals, Radnor, US) and deionized water at various volume ratios to tune the solvent viscosity. The mixtures are denoted as GxWy, where G and W represent the volume fractions (in %) of glycerol and water, respectively (e.g., G100W0, G70W30, and G50W50). Customized monodisperse polystyrene particles were then added at prescribed mass-to-volume concentrations, $c$% (w/v), defined as $c$ g of particles per 100 mL of the solvent mixture. After the addition of particles, the solution was heated in a 70 °C water bath for 30 minutes and subsequently subjected to ultrasonic stirring for 1 hour to ensure complete homogenization and particle dispersion, resulting in the final sample preparation.

**Data Description**

Based on this setup, our dataset includes two observables derived from each recorded speckle sequence: $g_2(t)$ and the speckle intensity histogram. The former provides the core dynamic information, while the latter provides the static context necessary to resolve optical ambiguity.

**Autocorrelation curves: Carrier of dynamic information**

When the scattered light from the sample is collected by a camera, an interference pattern, known as a speckle pattern, can be observed [39], see Figure 4. Autocorrelation specifically measures similarity between speckle patterns over time and directly corresponds to the rate at which Brownian motion causes the scattered field to lose phase coherence. Here, the $g_2(t)$ curve can be written by [28]

$$g_2(t) = <\frac{\langle I(t_0)I(t_0+t)\rangle}{\sqrt{\langle I(t_0)^2\rangle\langle I(t_0+t)^2\rangle}}>_{t_0} \tag{8}$$

Where $I(t_0)$ and $I(t_0 + t)$ denote the speckle pattern intensity at the time $t_0$ and time of $t_0 + t$, and the term $\langle I(t_0)I(t_0 + t)\rangle$ is the cross-correlation expectation of these two speckle patterns.

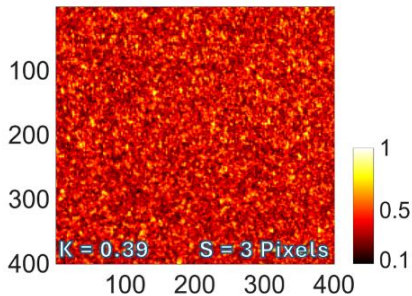

Figure 4 Typical speckle pattern. S is the speckle size, and K is the contrast of this speckle pattern. S is estimated as the spatial correlation length, taken as the Full Width at Half Maximum (FWHM) of the peak of the 2D intensity autocorrelation function (in pixels). The color bar indicates the normalized intensity. The K is calculated by the standard deviation of the speckle intensity divided by the mean intensity. The sidebar shows the normalized intensity level.

Due to this intensity dependence, the $g_2(t)$ curve is influenced not only by the sample dynamic feature but also by the optical parameters of the sample. Therefore, two samples with similar dynamics, but different optical conditions may produce different $g_2(t)$ decays, as shown in Figure 5(a). Conversely, changes in particle concentration can reshape $g_2(t)$, even if the underlying viscoelastic response remains unchanged. This coupling between optical structure and mechanical dynamics prevents $g_2(t)$ from serving as the core parameter that reliably tracks the sample state. Consequently, the use of only $g_2(t)$ limits the accuracy of the extracted $G'(\omega)$ and $G''(\omega)$. For this reason, we also consider the accompanying speckle intensity histogram, which is directly available from the recorded speckle frames.

**Histogram: Supplementing static information**

Previous studies, including our own, have shown that the histogram's shape is governed by the scattering path length diversity and the detection geometry [27, 30-34]. These properties are sensitive to the sample's optical coefficients (absorption and scattering) as well as instrumental parameters such as the numerical aperture, the pixel-speckle ratio, and the exposure time relative to the decorrelation time [29, 40-44]. Therefore, using histograms together with $g_2(t)$ provides orthogonal processing: $g_2(t)$ primarily encodes dynamics, time-dependent phase decorrelation, while the histogram summarizes static optical conditions mentioned above. This can help the model distinguish between dynamic and optical factors. In the learning framework, the histogram channel informs the network about path statistics and detection averages, so that the network can attribute changes in $g_2(t)$ to motion rather than changes in light transport, improving discriminability and robustness.

To show the complementary information captured by the autocorrelation function $g_2(t)$ and the intensity histogram, we measured polystyrene suspensions with different particle sizes (0.49, 5, and 11μm) at a constant concentration of 2% (w/v) under identical solvent conditions (G100W0). As shown in Figure 5, the $g_2(t)$ curves for different sizes can be similar, whereas the corresponding intensity histograms are well separated, providing an additional cue for distinguishing samples. Additional sample datasets not shown in the main figures are provided in the repository [45].

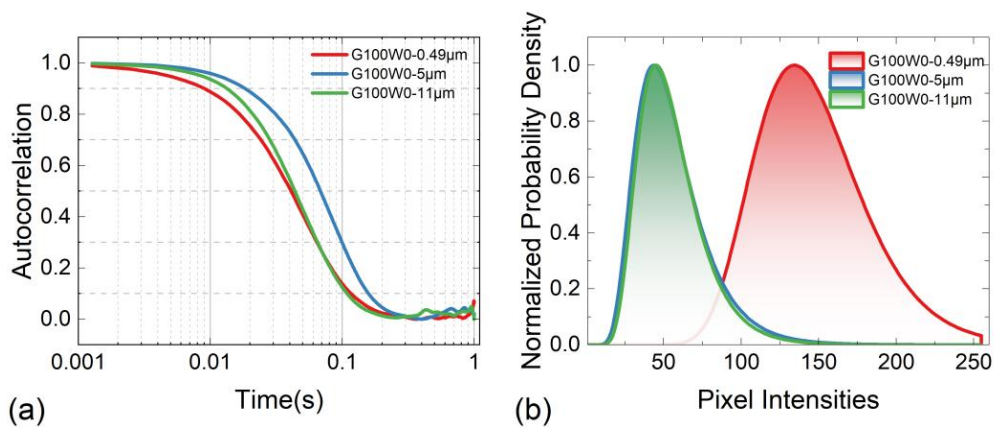

Figure 5 Complementarity of (a) $g_2(t)$ and intensity (b) histograms. Polystyrene suspensions in glycerol-water (100:0 w/w). (a, b) Varying particle size (0.49, 5, 11 μm) at constant concentration (2% w/v).

**Dataset generation and network training**

During the network training phase, we captured speckle patterns from samples with three different dynamic features: G50W50. G70W30, and G100W0. For each group of samples, the samples were prepared with polystyrene concentrations of 2%, 6%, and 12% and with particle sizes of 0.49, 1.7, and 5 µm to emulate practical sources of variation in scattering strength and mechanical response encountered in formulation and process settings. During the collection of speckle patterns, the sample temperature was precisely controlled at 25°C. The camera exposure time was regulated using the Pylon camera control software, with the region of interest (ROI) set to 100 × 100 pixels. The sampling rate was fixed at 790 frames per second, with an exposure time of 100 µs. Each recording lasted for 0.4 s to reduce redundant data, resulting in 316 consecutive frames per sample, which were treated as one sample's raw data segment. A sliding-time-window method was further applied to expand the dataset. The expanded dataset was split into training and testing sets, with 60% and 20% of the data used for training and validation, and the remaining 20% reserved for testing. To obtain the optimal reference for $G'(\omega)$ and ($G''(\omega)$) of the prepared sample, the advanced Anton Paar MCR 502e rheometer was used, with the temperature also maintained at 25°C during the measurement.

The training process was divided into two sequential stages. During the first stage, the network was optimized for 35 epochs to reconstruct the $g_2(t)$ curves and to capture the main decaying modes of such curves. In the second stage, an additional physics-based constraint was introduced, allowing the model to further align its prediction with the rheological properties of samples. Moreover, a mixup data augmentation strategy was incorporated during this stage to improve the network's generalization ability and robustness against overfitting [46]. The complete training spanned 300 epochs with a batch size of 32. Model selection was based on the minimum compound validation loss, and early stopping was applied to prevent overfitting. After the physics training was activated, the physics loss rapidly decreased and stabilized around $2.6\times10^{-2}$ (RMSE), while the $g_2(t)$ reconstruction error remained at the $10^{-4}$ level, shown in Figure S1. This two-stage strategy first ensures that the model captures discernible temporal modes, then refines them in the physical domain for precise alignment between prediction and experimental data. Detailed information can be found in the supplementary material.

## Results and Discussions

**Validation across test dataset**

As a first test, we examine whether the network can robustly reconstruct rheological responses under perturbed experimental conditions. Specifically, the test data in this section consisted of $g_2(t)$ signals and their corresponding histograms from samples with the same particle sizes and concentrations as those used in training, but collected at a different time and not included in the training set. This setting allows us to assess the network's stability and its ability to generalize to unseen measurement variations rather than to unseen material compositions.

As shown in Figure 6(a), the reconstructed $g_2(t)$ curves exhibit excellent alignment with the test data across all solvent ratios, achieving low root-mean-square errors, RMSE, (RMSEs ranging from 0.006 to 0.035) and indicating strong resilience to perturbations. The reconstructed relaxation spectra in Figure 6(b) are well-separated and physically interpretable. Each relaxation spectrum was reconstructed from speckle frames acquired within 0.4 s, corresponding to an equivalent frequency range of approximately 2.5-790 rad·s$^{-1}$. The model automatically decomposed the response into two dominant Maxwell modes. Importantly, the extracted relaxation times satisfy the expected trend with solvent composition: as the water fraction increases, the characteristic relaxation times systematically decrease, consistent with faster structural relaxation in less-viscous environments.

The resulting viscoelastic spectra $G'(\omega)$ and $(G''(\omega)$ in Figure 6(c) accurately reproduce both the amplitude and scaling trends. To ensure a robust quantitative comparison, we evaluated prediction errors against a smoothed reference baseline, strategically focusing on the high-frequency range (10-100 rad/s) where rheometer data is most reliable for low viscosity samples. We employed two key metrics: the logarithmic RMSE (RMSE$_{log}$), reflecting the average multiplicative error, and the Median Absolute Percentage Error (MdAPE), a robust measure of the typical relative error. For the seen and untrained sample, the model demonstrates excellent performance. The $G''(\omega)$ prediction is exceptionally accurate, with an RMSE$_{log}$ as low as 0.009 and a typical MdAPE of under 2.4%. The $G'(\omega)$ predictions are also strong, with best-case RMSE$_{log}$ values around 0.023-0.034 and a typical MdAPE of 5-7%. These results confirm that the recovered Maxwell modes are not only mathematically accurate but also remain quantitatively consistent with the underlying physical dependence of relaxation time on solvent viscosity, demonstrating the network's high reconstruction fidelity

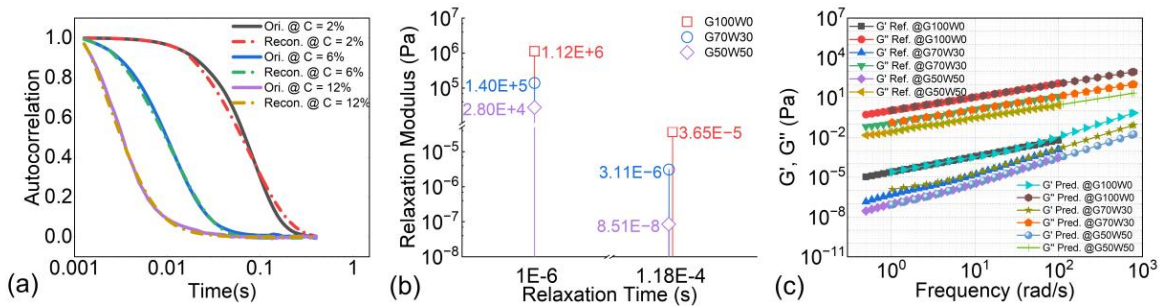

Figure 6 (a) Reconstructed and experimental $g_2(t)$ curves showing excellent agreement across trained glycerol-water ratios (G100W0, G70W30, and G50W50). (b) Corresponding relaxation spectra recovered from speckle frames, decomposed into three dominant Maxwell modes. (c) Predicted $G'(\omega)$ and ($G''(\omega)$ spectra maintain correct scaling trends and consistent viscosity-dependent relaxation behavior.

**Evaluation on Unprecedented Material Conditions**

To evaluate further the model's generalizability, we tested three unprecedented conditions in the same glycerol-water system (G70W30): varying particle size, varying concentration, and varying both simultaneously. The primary challenge in analyzing such highly concentrated suspensions lies in the multiple scattering effect, which significantly distorts the autocorrelation function. Figures 7(a, d, g) demonstrate that the network reconstructs the $g_2(t)$ curves with high fidelity. This visual alignment is confirmed by low $g_2(t)$ reconstruction RMSE values, typically between 0.004 and 0.032 across all test cases, accurately recovering the intrinsic decay dynamics of the system. The Maxwell relaxation spectra derived from these reconstructed curves (Figures 7(b, e, h)) further demonstrate the model's physical consistency. Importantly, the model accurately distinguishes between the physical effects of particle size and those of concentration. For the case where only particle size is varied, the experimentally measured $G'(\omega)$ and $G''(\omega)$ remain virtually unchanged, and the model predicts Maxwell modes with highly consistent relaxation times ($\tau$), reflecting the substantial invariance of the solvent viscosity and fluid dynamics. In contrast, for tests with varying concentration, the model identifies the primary differences as arising from changes in the relaxation modulus $G$, while the relaxation time distribution remains virtually unchanged. This is consistent with physical laws:

relaxation time is primarily determined by the solvent environment, while particle concentration primarily controls the system's modulus.

In the final viscoelastic spectrum predictions (Figure 7(c, f, i)), the model performed most stably in predicting the loss modulus $G''(\omega)$. The predictions for $G''(\omega)$ achieved a low typical MdAPE below 15% and a favorable $RMSE_{log}$ that ranged from 0.017 to 0.075 across the test cases. Although the storage modulus $G'(\omega)$ exhibited a larger relative error, with MdAPE values ranging from 41% to 69% and an $RMSE_{log}$ between 0.16 and 0.24 for unseen concentration cases, this can be attributed to its inherently small signal amplitude (typically several orders of magnitude lower than $G''(\omega)$), where even small absolute deviations can lead to large relative errors. More importantly, the model accurately reproduced the log-log scale slope and relative hierarchy of $G'(\omega)$ and $G''(\omega)$ across all samples, demonstrating its exceptional ability to learn low-signal physics, maintain high quantitative accuracy, and separate particle size and concentration effects. These results demonstrate that the model not only effectively eliminates the effects of multiple scattering but also extracts core dynamical information from the raw speckle images, enabling physically consistent and interpretable viscoelastic spectrum reconstruction.

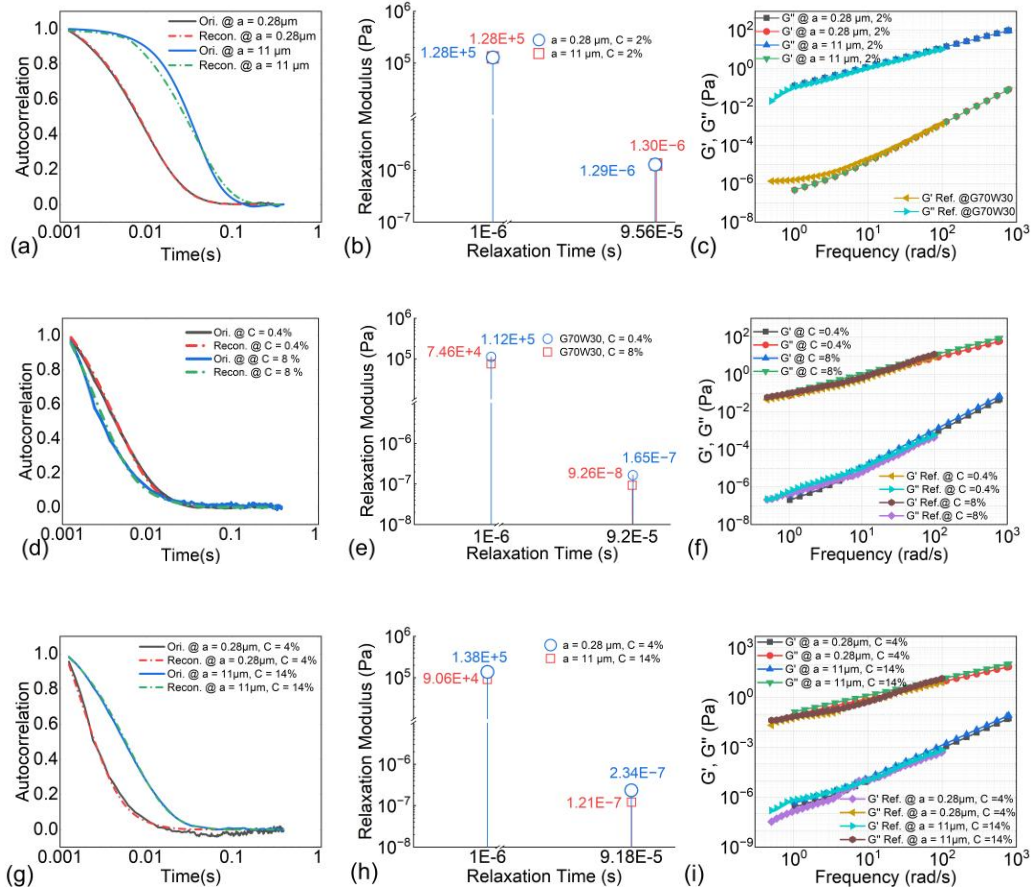

Figure 7. Generalization performance of the proposed network on unseen particle sizes and concentrations in glycerol-water suspensions (G70W30). (a-c) Tests with unseen particle sizes (0.28 μm and 11 μm, 2%(w/v)). (d-f) Tests with unseen concentrations (0.4%(w/v) and 8%(w/v), 5 μm). (g-i) Tests with both unseen particle size (0.28 μm) and concentration (14%(w/v)). The network reconstructs single-scattering $g_2(t)$ curves with high fidelity (a, d, g), yielding physically consistent Maxwell spectra (b, e, h) and viscoelastic moduli $G'(\omega)$ and ($G''(\omega)$) (c, f, i). The relaxation times remain nearly constant across conditions, while modulus magnitudes vary with concentration, consistent with the expected dependence on particle loading rather than solvent dynamics.

Furthermore, we designed two sets of extreme tests (Figure 8) to further test the model's limits. The first set of samples used different glycerol-water ratios (G60W40 and G80W20) from the training set, while maintaining the same particle size and concentration (5 μm, 2% ((w/v))). This was used to evaluate the consistency of the response under varying solvent scattering conditions. The second set of samples simultaneously varied the solvent, particle size, and concentration (11 μm, 3%(w/v) and 0.28 μm, 14%(w/v)), representing combinations not encountered during training. This allowed us to test the model's generalization limits under unseen dynamics.

The results show that even under conditions outside the training range, the network is able to reconstruct reasonable $g_2(t)$ curves with low reconstruction RMSEs, typically between 0.008 and 0.02, and generate physically interpretable Maxwell spectra. Figures 8(b) and 6(e) show systematic shifts in the principal relaxation times of the G60W40 and G80W20 samples, reflecting kinetic differences caused by changes in solution viscosity and refractive index. This trend demonstrates that the model can distinguish between relaxation changes caused by changes in system viscosity and signal distortions due to differences in scattering paths. Quantitatively, for these two test sets, the MdAPE for $G'(\omega)$ fluctuates between 27-78% and the RMSE$_{log}$ ranges from 0.08 to 0.40, while the MdAPE for $(G''(\omega))$ ranges from approximately 20-55% with a corresponding RMSE$_{log}$ between 0.10 and 0.35. Despite relatively higher MdAPE and RMSE$_{log}$ values compared to those observed for training-range data, the predicted results maintain the correct relationship between the frequency-dependent slope and the relative scaling of the modulus for all samples. This "large error but correct pattern" phenomenon demonstrates that the model adheres strictly to physical constraints under completely unseen conditions, rather than relying on a black-box numerical fit.

The Maxwell spectra show that the model can reliably identify the dominant relaxation modes in different scattering environments. These results suggest that the network captures distinct dynamic relaxation and static scattering contributions by integrating speckle histogram statistics with $g_2(t)$ decay characteristics. Notably, for completely unseen experimental conditions, the predicted relaxation time interval, modulus ratio, and frequency dependence remain within physically reasonable ranges. Taken together, these observations are consistent with physics-informed learning behavior and indicate good generalization capability.

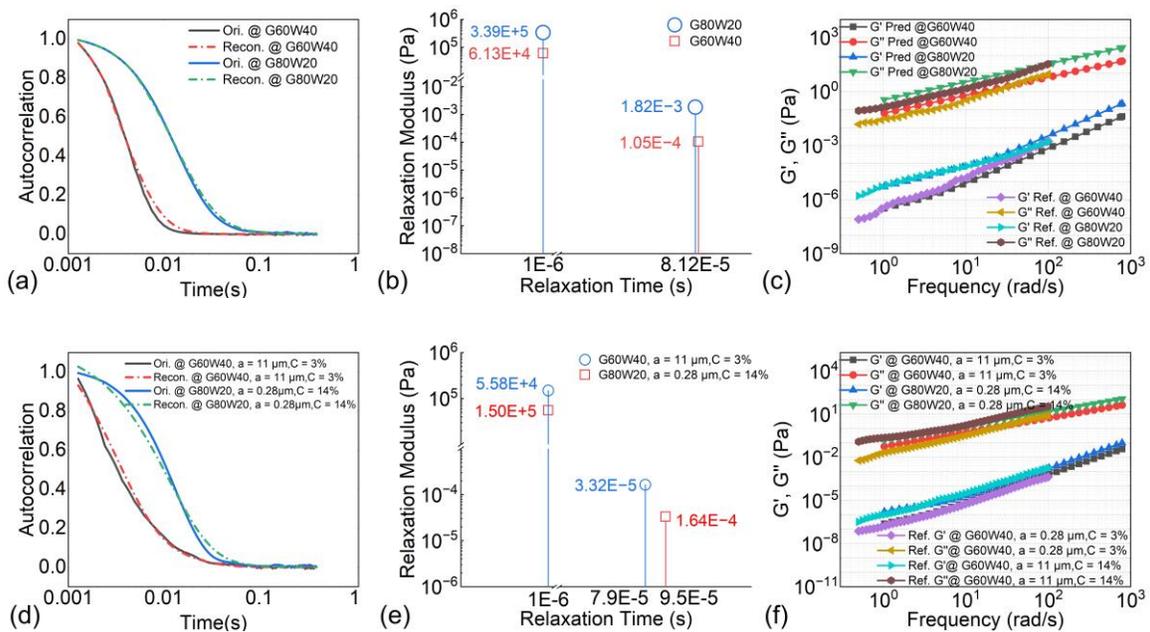

Figure 8. (a-c) Results for samples with unseen solvent compositions (G60W40 and G80W20) but trained on particle size and concentration. (d-f) Results for fully unseen combinations of solvent, particle size, and concentration. Despite higher numerical errors in magnitude, the predicted relaxation behavior remains physically valid, indicating that the network separates dynamic relaxation from scattering-induced variations rather than performing empirical fitting.

**Ablation test**

The physical significance and value of including the histogram information were further verified through ablation testing (Figure 9). We compared the predictions of the full model with those after removing the histogram. We used a system with a particle size of 5 μm and a concentration of 2%(w/v) in three typical solvents (G50W50, G70W30, and G100W0). The results showed that when the histogram was removed and only $g_2(t)$ was used for viscoelastic spectrum prediction, both $G'(\omega)$ and $(G''(\omega)$ were systematically overstated. This trend is consistent with the results obtained when solving the generalized Stokes-Einstein equations (GSER) using the traditional uncalibrated MSD-$g_2(t)$ relationship, indicating that the histogram serves as a statistical scale normalization and optical correction factor in the network.

From a physical perspective, the histogram branch provides an "optical flux" constraint directly related to speckle statistics, enabling the network to distinguish between dynamic relaxation (particle Brownian motion) and static scattering (light path distribution), thereby achieving physically consistent predictions rather than empirical fits. This branch also has extremely low computational complexity, processing only the statistical histogram rather than the raw image sequence. Compared to end-to-end image input, this design significantly reduces computational and storage costs while maintaining prediction accuracy, resulting in higher real-time performance and scalability, making the model more relevant to practical industrial applications.

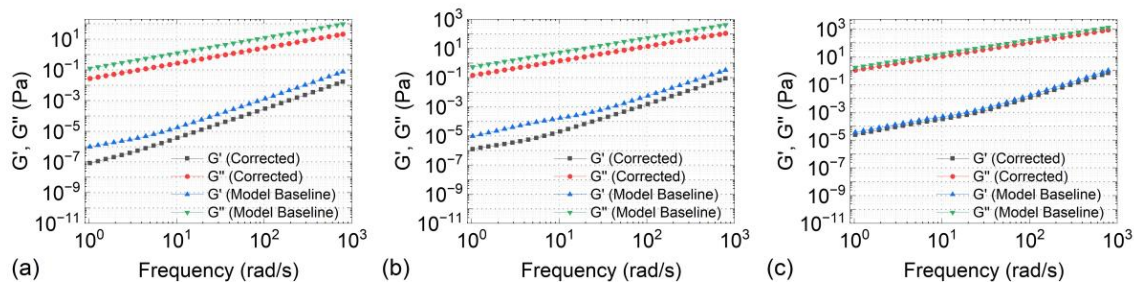

Figure 9. Results for 5 μm particles (2 % (w/v)) in G50W50 (a), G70W30(b), and G100W0(c) solvents. Without the histogram input, both $G'(\omega)$ and ($G''(\omega)$ are systematically overestimated.

**Stability test**

To assess the long-term reliability, which is crucial for practical applications, we evaluated the network's predictive stability over a 10-hour. We acquired speckle data hourly from challenging, previously unseen samples (11 μm particles, 3% w/v, G80W20 and G60W40). We quantified drift by comparing each subsequent prediction with the initial (t=0) prediction using RMSE$_{log}$ and MdAPE. As shown in Figure 10, the network exhibits strong stability. The prediction errors for both $G'(\omega)$ and $G''(\omega)$ remain low, especially in the initial four hours: MdAPE for $G'(\omega)$ and ($G''(\omega)$ is below 0.2% and RMSE$_{log}$ below 0.001 in G60W40, while in G60W40, MdAPE for $G'(\omega)$ and ($G''(\omega)$ is below 3% and RMSE$_{log}$ below 0.03. Although the error metrics rise slightly over the entire 10 hours, this slight drift is likely due to accumulated experimental factors such as system noise and potential particle settling, rather than model instability. Crucially, no bias emerges, and the errors remain consistently within manageable limits. Ultimately, the MdAPE for all $G'(\omega)$ and ($G''(\omega)$ is below 1% and 8%, respectively. This consistent fidelity highlights the network's robustness in maintaining stable performance over extended periods relevant to process monitoring. This consistent and sustained high fidelity, even in the face of prolonged measurement challenges and potential sample evolution, strongly underscores the network's inherent robustness and validates its potential for reliable deployment in applications like real-time process monitoring where consistent performance over extended durations is paramount.

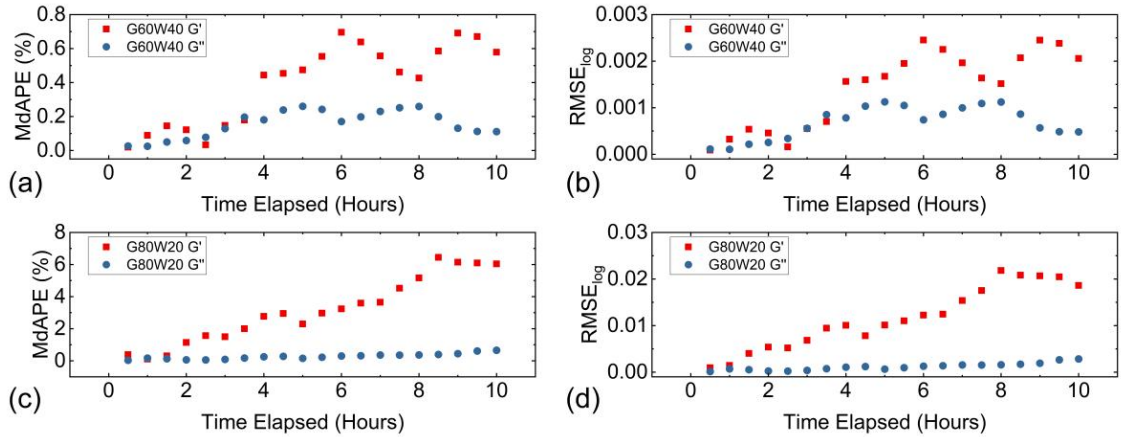

Figure 10 Long-term stability of network predictions over a 10-hour for unseen samples (11 μm particles, 3% in G80W20 and G60W40). Subplots show MdAPE (%) and RMSE$_{log}$ drift for $G'(\omega)$ and $(G''(\omega)$ relative to the initial prediction (t=0). The contained error demonstrates model robustness for continuous monitoring.

## Conclusion

This study proposes a framework that integrates physical constraints and deep learning to directly reconstruct the storage modulus $G'(\omega)$ and loss modulus $(G''(\omega)$ from multi-scattering speckle patterns. The overall network design follows a physically interpretable structure: a) the decay analyzer module extracts time decay features and interprets their multiscale relaxation behavior; b) the initial Maxwell module generates a physically interpretable initial modulus spectrum using a parameterized distribution; c) attention implements weighted coupling between dynamic and static features; d) finally, the physics forward layer completes physically consistent frequency-domain mapping. This layered structure explicitly embeds physical laws into the network flow, enabling interpretable modeling throughout the entire chain, from signal analysis and feature fusion to physical output. This design enables the network to directly extract physical information reflecting particle Brownian dynamics from experimental speckle patterns without relying on any explicit scattering model.

Systematic experiments demonstrated that the model reliably predicted physically consistent viscoelastic spectra across a wide range of test conditions. For samples with known solvents but no known particle size or concentration, the MdAPE of $G''(\omega)$ was typically less than 15%. While the error for $G'(\omega)$ was slightly larger (20-60%), it maintained the correct frequency dependence and scaling trend. In completely unseen solvent systems (G60W40 and G80W20), the model still accurately identified the main relaxation time distributions, and the Maxwell mode migration direction was consistent with the viscosity trend, demonstrating that its generalization ability stems from true learning of dynamical and optical statistics rather than empirical fitting. Furthermore, the histogram branch significantly improved prediction stability and reduced computational overhead, resulting in an approximately order-of-magnitude reduction in overall computational cost compared to end-to-end networks directly based on image sequences, demonstrating the potential for industrial-scale, real-time characterization.

Nevertheless, the model's performance remains limited by the coverage of the training data, particularly in extreme optical conditions and complex dispersions. Future work will focus on developing more fine-grained training sample generation strategies and exploring the integration of additional modal information (such as spectral, phase, and polarization signals) to provide more comprehensive physical constraints and characterization. This multimodal physics-driven network is not only expected to further improve prediction accuracy and robustness, but also lays a theoretical and technical foundation for the intelligent application of speckle imaging in non-invasive rheological measurements and the characterization of complex fluids.

## Acknowledgement

This IGF project of the FEI was supported within the programme for promoting the Industrial Collective Research (IGF) of the German Federal Ministry for Economic Affairs and Energy (BMWE) on the basis of a decision by the German Bundestag. Project 01IF22490N. This work was also supported by the Engineering and Physical Sciences Research Council (EPSRC), UK, under grant number EP/Z53285X/1. Open Access funding enabled and organized by Projekt DEAL. The authors acknowledge Applied Microspheres GmbH for providing custom-sized polystyrene microspheres.

## Authors' contributions

T.W. conceived the concept, carried out the theoretical analysis, developed the algorithms and optimizations, conducted the experiments, analyzed, interpreted the data, and drafted the manuscript. T.G. contributed to the experimental investigations, participated in result interpretation and discussions, and co-wrote the manuscript. E.F., Y.Y., and I.E. provided algorithmic support and contributed to the development and validation of the computational methods. D.G., Y.Liu, Y.Li, and M.B. contributed to result interpretation and discussions. E.F.,

D.G., and T.B. provided financial support for the project. All authors reviewed, edited, and approved the final manuscript.

## Data availability

The data that support the findings of this study are available from the corresponding author upon reasonable request.

The datasets showing the complementary during the current study are available in the Zenodo repository, https://doi.org/10.5281/zenodo.18186797 [44]

## Declarations

### Competing interests

The authors declare no conflict of interest.

## Reference

1. Català-Castro, F., et al., *Measuring age-dependent viscoelasticity of organelles, cells and organisms with time-shared optical tweezer microrheology.* Nature Nanotechnology, 2025. **20**(3): p. 411–420.

2. Conley, G.M., et al., *Relationship between rheology and structure of interpenetrating, deforming and compressing microgels.* Nature communications, 2019. **10**(1): p. 2436.

3. Khondkar, D., et al., *Rheological behaviour of uncross-linked and cross-linked gelatinised waxy maize starch with pectin gels*. Food Hydrocolloids, 2007. **21**(8): p. 1296–1301.

4. Khosravi, A., et al., *Understanding the rheology of nanocontacts*. Nature Communications, 2022. **13**(1): p. 2428.

5. Jiménez-Avalos, H., E. Ramos-Ramírez, and J. Salazar-Montoya, *Viscoelastic characterization of gum arabic and maize starch mixture using the Maxwell model*. Carbohydrate Polymers, 2005. **62**(1): p. 11–18.

6. Lu, Y.-B., et al., *Viscoelastic properties of individual glial cells and neurons in the CNS*. Proceedings of the National Academy of Sciences, 2006. **103**(47): p. 17759–17764.

7. Šćepanović, P., T.B. Goudoulas, and N. Germann, *Numerical investigation of microstructural damage during kneading of wheat dough*. Food structure, 2018. **16**: p. 8–16.

8. Besiri, I.N., et al., *In situ evaluation of alginate –Ca2+ gelation kinetics*. Journal of Applied Polymer Science, 2023. **140**(32): p. e54252.

9. Malkin, A.Y. and A.I. Isayev, *Rheology: concepts, methods, and applications*. 2022: Elsevier.

10. Singh, P.K., J.M. Soulages, and R.H. Ewoldt, *Frequency-sweep medium-amplitude oscillatory shear (MAOS)*. Journal of Rheology, 2018. **62**(1): p. 277–293.

11. Laun, M., et al., *Guidelines for checking performance and verifying accuracy of rotational rheometers: viscosity measurements in steady and oscillatory shear (IUPAC Technical Report)*. Pure and Applied Chemistry, 2014. **86**(12): p. 1945–1968.

12. Del Giudice, F., *A review of microfluidic devices for rheological characterisation.* Micromachines, 2022. **13**(2): p. 167.

13. Hyun, K., et al., *A review of nonlinear oscillatory shear tests: Analysis and application of large amplitude oscillatory shear (LAOS).* Progress in polymer science, 2011. **36**(12): p. 1697–1753.

14. Ricarte, R.G. and S. Shanbhag, *A tutorial review of linear rheology for polymer chemists: basics and best practices for covalent adaptable networks.* Polymer Chemistry, 2024. **15**(9): p. 815–846.

15. Hnyluchová, Z., et al., *A simple microviscometric approach based on Brownian motion tracking.* Review of Scientific Instruments, 2015. **86**(2).

16. Mao, Y., P. Nielsen, and J. Ali, *Passive and active microrheology for biomedical systems.* Frontiers in bioengineering and biotechnology, 2022. **10**: p. 916354.

17. Läuger, J. and H. Stettin, *Effects of instrument and fluid inertia in oscillatory shear in rotational rheometers.* Journal of Rheology, 2016. **60**(3): p. 393–406.

18. Stetefeld, J., S.A. McKenna, and T.R. Patel, *Dynamic light scattering: a practical guide and applications in biomedical sciences.* Biophysical reviews, 2016. **8**(4): p. 409–427.

19. Medebach, M., et al., *Dynamic light scattering in turbid colloidal dispersions: A comparison between the modified flat-cell light-scattering instrument and 3D dynamic light-scattering instrument.* Journal of colloid and interface science, 2007. **305**(1): p. 88–93.

20. Pine, D.J., et al. *Features of diffusing wave spectroscopy*. in *Photon Correlation Techniques and Applications*. 1988. Optica Publishing Group.

21. Weitz, D.A. and D.J. Pine, *Diffusing-wave spectroscopy.* Dynamic light scattering, 1993: p. 652–720.

22. Zhang, C., et al., *Improved diffusing wave spectroscopy based on the automatized determination of the optical transport and absorption mean free path.* Korea-Australia Rheology Journal, 2017. **29**(4): p. 241–247.

23. Fahimi, Z., et al., *Diffusing-wave spectroscopy in a standard dynamic light scattering setup.* Physical Review E, 2017. **96**(6): p. 062611.

24. Kim, H.S., et al., *Diffusing wave microrheology of highly scattering concentrated monodisperse emulsions.* Proceedings of the National Academy of Sciences, 2019. **116**(16): p. 7766–7771.

25. Mason, T.G. and D.A. Weitz, *Optical measurements of frequency-dependent linear viscoelastic moduli of complex fluids.* Physical review letters, 1995. **74**(7): p. 1250.

26. Hajjarian, Z. and S.K. Nadkarni, *Tutorial on laser speckle rheology: technology, applications, and opportunities.* Journal of biomedical optics, 2020. **25**(5): p. 050801–050801.

27. Hajjarian, Z. and S.K. Nadkarni, *Evaluation and correction for optical scattering variations in laser speckle rheology of biological fluids.* PloS one, 2013. **8**(5): p. e65014.

28. Leartprapun, N., et al., *Laser speckle rheological microscopy reveals wideband viscoelastic spectra of biological tissues.* Science Advances, 2024. **10**(19): p. eadl1586.

29.	Postnov, D.D., et al., *Dynamic light scattering imaging*. Science advances, 2020. **6**(45): p. eabc4628.

30.	Michau, G., G. Frusque, and O. Fink, *Fully learnable deep wavelet transform for unsupervised monitoring of high-frequency time series*. Proceedings of the National Academy of Sciences, 2022. **119**(8): p. e2106598119.

31.	Hu, J., L. Shen, and G. Sun. *Squeeze-and-excitation networks*. in *Proceedings of the IEEE conference on computer vision and pattern recognition*. 2018.

32.	Kingma, D.P. and M. Welling, *Auto-encoding variational bayes*. arXiv preprint arXiv:1312.6114, 2013.

33.	Rezende, D. and S. Mohamed. *Variational inference with normalizing flows*. in *International conference on machine learning*. 2015. PMLR.

34.	Baumgaertel, M. and H.H. Winter, *Determination of discrete relaxation and retardation time spectra from dynamic mechanical data*. Rheologica Acta, 1989. **28**(6): p. 511–519.

35.	Vaswani, A., et al., *Attention is all you need*. Advances in neural information processing systems, 2017. **30**.

36.	Dosovitskiy, A., *An image is worth 16x16 words: Transformers for image recognition at scale*. arXiv preprint arXiv:2010.11929, 2020.

37.	Ferry, J.D., *Viscoelastic properties of polymers*. 1980: John Wiley & Sons.

38.	Sutton, M., et al., *Observation of speckle by diffraction with coherent X-rays*. Nature, 1991. **352**(6336): p. 608–610.

39. Goodman, J.W., *Speckle phenomena in optics: theory and applications*. 2007, Roberts and Company Publishers.

40. Wang, T., et al., *Optimized Laser Speckle Rheology Measurement Based on Speckle Pattern's Gamma Correction and Neural Network*. Optics & Laser Technology, 2025. **191**: p. 113320.

41. Goodman, J.W., *Some fundamental properties of speckle*. JOSA, 1976. **66**(11): p. 1145–1150.

42. Cummins, H., *Photon correlation and light beating spectroscopy*. Vol. 3. 2013: Springer Science & Business Media.

43. Majic, M., W.R. Somerville, and E.C. Le Ru, *Path-length distributions, scattering, and absorption in refractive spheres and slabs*. Physical Review A, 2023. **107**(5): p. 053509.

44. Ramirez-San-Juan, J., et al., *Effects of speckle/pixel size ratio on temporal and spatial speckle-contrast analysis of dynamic scattering systems: Implications for measurements of blood-flow dynamics*. Biomedical optics express, 2013. **4**(10): p. 1883–1889.

45. Wang, T., *Raw laser speckle rheology dataset: g2 curves and intensity histograms for G50/G70/G100 suspensions (0.49, 5, 11 μm)*. 1.0 ed, ed. Zenodo. 2026.

46. Thulasidasan, S., et al., *On mixup training: Improved calibration and predictive uncertainty for deep neural networks*. Advances in neural information processing systems, 2019. **32**.